\documentclass[aps,showpacs,pre,twocolumn,groupedaddress]{revtex4}
\usepackage{amsmath,amsfonts,amssymb}

\usepackage{graphicx,color}
\begin{document}

\title{Velocity-correlation distributions in granular systems}

\author{Alexis Burdeau}
\author{Pascal Viot}

\affiliation{Laboratoire de Physique
Th\'eorique de la Mati\`ere Condens\'ee, UMR 7600, Universit\'e Pierre et Marie
 Curie, 4, place
Jussieu, 75252 Paris Cedex 05, France}

\bibliographystyle{prsty}

\begin{abstract}
We  investigate the   velocity-correlation  distributions   after  $n$
collisions  of   a   tagged  particle undergoing   binary  collisions.
Analytical expressions are obtained in any dimension  for the
velocity-correlation distribution after the first-collision as well as
for  the velocity-correlation function  after   an infinite number  of
collisions, in   the  limit of Gaussian   velocity  distributions.  It
appears that  the decay  of  the first-collision  velocity-correlation
distribution for negative  argument is  exponential in any  dimension
with a coefficient that depends on the mass and on the coefficient of
restitution.  We also  obtained the  velocity-correlation distribution
when the velocity distributions are not  Gaussian: by inserting Sonine
corrections of the  velocity distributions, we derive the corrections to
the velocity-correlation distribution which agree perfectly with  a
DSMC (Direct Simulation Monte Carlo) simulation.  We emphasize that these new quantities can be easily
obtained in simulations and likely   in experiments: they could be  an
efficient probe    of  the local environment   and   of the  degree of
inelasticity of the collisions.
\end{abstract}

\pacs{05.20.Dd, 51.10.+y, 02.50.-r}

\maketitle

\section{Introduction}
The dynamics of hard-core   particles  consists    of successive binary
collisions.  For atomic systems, the equilibrium  state can be reached, and
is characterized  by velocity     distributions which are    purely
Gaussian.  Conversely, in   the presence  of  dissipation,  i.e.   for
granular particles, no equilibrium exists, but when an external
source of energy is present the system reaches  a steady state whose  properties
can   be    compared     to   the  equilibrium    state   of    atomic
systems\cite{BP004}.  At  low   to intermediate   densities,   spatial
correlations      are   not       responsible     for     non-Gaussian
deviations\cite{Baxter2007}.

The short-time dynamics  is usually   analyzed  by means of   the velocity
autocorrelation function. This quantity provides an average of the  scalar product of  the velocity at  time $0$  with the
velocity  at  time $t$.   The  characteristic time of this correlation
function corresponds to the   time needed for the system to
lose memory  of the initial  configuration of the  velocities.  Some
progress has been  made  recently  by investigating  the   collision
statistics: Visco  {\it et. al}\cite{Visco2008,Visco2008a} showed that
the  free  flight time  distribution  is not exponential, even  in the
low-density limit (such a behavior was  observed in Molecular Simulation
of hard spheres some years ago\cite{Talbot1992}). Deviations from the
Poisson  law  of the  number  of collisions  can  be  captured in  the
framework of  the   Boltzmann equation  and  agree   with  molecular
simulation results.

We introduce   here a new  quantity  by focusing  on  collision events
irrespective of  the time when individual collisions  occur. We
consider   the scalar product  between  the  velocity  before  a given
collision and  the   velocity after $n$  collisions.  (Note   that this
quantity is distinct from the distribution of velocity for a hard sphere
on collision, which   characterizes the distribution  of  the relative
velocity of colliding  spheres, for which Lue\cite{Lue2005} obtained exact
results in three, four and  five dimensions).  The information obtained is not only
the  average of the  scalar product, but the
full distribution of the scalar
product between the velocity  before and after a sequence of $n$ collisions.  When $n=1$, this corresponds to the probability
of the scalar product between the pre- and post-collisional velocities
during  a collision; it  is worth noting  that the first moment of the
distribution does not  correspond to the velocity correlation function
at  the mean collision  time: indeed,  the  probability distribution is
built for collisions occurring  at  different collision times,  whereas
the velocity   correlation function   corresponds  to the velocities scalar product at a given time.

As the number    of collisions increases,   the correlations between
velocities decrease and the distribution  evolves progressively to the
asymptotic  form where the velocities  are uncorrelated.  The paper is
organized as follows:     in  section  II,   we  obtain  the  probability
distributions  in the   limit  of an infinite  number of   collisions, i.e. when the  velocities  are completely uncorrelated, in  any
spatial dimension. In section III, we  derive  the first-collision velocity
distribution in   any   dimension. Section  IV is devoted to the
corrections induced by the  non-Gaussian behavior of  granular systems,
and we  compare the analytical results to DSMC results. 
Velocity-correlation distributions at the second and higher collisions are analyzed in Section V.

Let  us define, right away, the central quantity of this study, the velocity-correlation   distribution $P_n(z)$ at the
$n$th collision :
\begin{equation}\label{eq:1}
P_n(z)=\langle \delta(z-{\bf v}.{\bf v}^*_{n})\rangle
\end{equation}
where  the brackets  denote a statistical  average in a
given steady state, ${\bf v}$ denotes the  precollisional velocity of a
tagged  particle before  the first  collision   and ${\bf v}^*_n$  the
postcollisional velocity of   the  same    particle  after the     $n$th
collision. We only consider the case of homogeneous systems.

\section{Velocity-correlation distribution in the infinite collision limit}
We first consider the  situation where the number  of collisions is 
very  large, i.e.  the  velocity  before  the  first collision and  the
velocity after a large number  of collisions become uncorrelated.  The
probability  distribution of  the  scalar product   $P_{\infty}(z={\bf v}.{\bf
v^*}_\infty)$ is then given by
\begin{equation}\label{eq:2}
P_{\infty}(z)=\int\int   d{\bf   v}    d{\bf    v^*_{\infty}}   f({\bf     v})f({\bf
v^*}_\infty)\delta(z-{\bf v}.{\bf v^*}_\infty).
\end{equation}

Let  us  consider    the generating  function   $\tilde{P}_{\infty}(k)=\int  dk
e^{ikz}P_{\infty}(z)$. One has
\begin{equation}\label{eq:3}
\tilde{P}^{(d)}_{\infty}(k)=\int\int   d{\bf   v}    d{\bf    v^*}_\infty   f({\bf     v})f({\bf
v^*}_\infty)e^{ik{\bf   v}.{\bf   v^*}_\infty}.
\end{equation}
We assume  that  the velocity  distribution  can be factorized  as
$f({\bf v})=\prod_{\alpha=1}^d  f(v_\alpha)$ where $\alpha$ is an index
running    over all components  of  the   velocity  and  $d$ the space
dimension. The Cartesian components of   the velocity are  independent
random variables and the generating function is then also  the product
of the generating functions of each Cartesian component:
\begin{equation}\label{eq:4}
\tilde{P}_{\infty}^{(d)}(k)=(\tilde{P}^{(1)}_\infty(k))^d,
\end{equation}
where $\tilde{P^{(1)}_\infty}(k)$ is the generating function for the  one-dimensional problem.

$P^{(1)}_\infty(z)$  can   be  obtained from
Eq.~(\ref{eq:2}), which gives
\begin{equation}
P^{(1)}_\infty(z)=\int \frac{dv}{|v|}f(z/v)f(v).
\end{equation}
It is worth noting  that ${P}^{(1)}_\infty$ is the Mellin  convolution of  the two
velocity  distributions as this distribution is that  of  the product of two  independent
random variables.  When the velocity distribution $f(v)$ is Gaussian
\begin{equation}
f(v)=\sqrt{\frac{M}{2\pi T}}e^{-\frac{Mv^2}{2T}},
\end{equation}
where $M$ and $T$ are the mass and the temperature of the tagged particle, respectively.

$P^{(1)}_\infty(z)$ can be explicitly obtained  and is then equal to
\begin{equation}\label{eq:5}
P^{(1)}_\infty(z)=\frac{MK_0(\frac{|z|M}{T})}{\pi T},
\end{equation}
where $K_0(z)$ is the modified Bessel function of second kind.

The basic property of ${P}^{(1)}_\infty$ is that the distribution is symmetric because  the velocities are   uncorrelated.  The behavior  of
${P}^{(1)}_\infty$ is intriguing at small values of $z$, as one
observes a logarithmic divergence at $z=0$.  This means that there is an overpopulated density  of  very
small scalar products even though the velocity distribution remains finite
for  very small  velocities.  For  large velocities,  ${P}^{(1)}_\infty$ decays as
$\exp(-|z|M/T)/\sqrt{z}$ for large   values of $z$, i.e. less   rapidly
than the original velocity distribution which has a Gaussian decay.

In two and more dimensions, the velocity-correlation distribution can be obtained by noting that 
 the  Fourier transform of Eq.(\ref{eq:5}) (or  integrating  Eq.(\ref{eq:3})), leads to 
\begin{equation}\label{eq:6}
\tilde{P}^{(1)}_\infty(k)=\frac{1}{\sqrt{1+k^2}}.
\end{equation}

By inserting    Eq.(\ref{eq:6}) in   Eq.(\ref{eq:4}),   the generating
function $\tilde{P}^{(d)}_\infty(k)$ in $d$ dimensions is then
\begin{equation}
\tilde{P}^{(d)}_\infty(k)=(1+k^2)^{-d/2}.
\end{equation}
The inverse Fourier transform can be  calculated in any dimension: in
2D, the Fourier transform has a Lorentz profile, which gives in real space
\begin{equation}
P^{(2)}_\infty(z)=\frac{M\exp(-|z|M/T)}{2T},
\end{equation}
and in three dimensions, 
\begin{equation}
P_\infty^{(3)}(z)=z\frac{MK_1(\frac{|z|M}{T})}{\pi T}.
\end{equation}
In 2D and in 3D, the probability distribution is no longer singular at the
origin. However, there exists a non-analytic behavior which is   a  cusp in 2D and  a cusp in the derivative in 3D.

For completeness, the solution in odd dimensions; $P^{(d)}_\infty(z)$ is given by
\begin{equation}
P^{(d)}_\infty(z)=\sqrt{\frac{2}{\pi}}|z|^{(d-1)/2}\frac{(d-2)!}{2^{(d-3)/2}((d-3)/2)!}K((d-1)/2,|z|)
\end{equation}collision
where $K((d-1)/2,|z|)$ is the modified Bessel function of second kind of order $(d-1)/2$.

\section{First-collision velocity-correlation distribution}
In order to have tractable expressions for the first-collision velocity-correlation
distribution  $P_1(z)$,  we assume  that  the the ``molecular chaos'' is
valid, i.e.  that there are no correlations between the particles before
collision. The joint velocity distribution of the tagged particle
and the  bath   particles is  simply the  product   of  the individual
velocity distributions.  Moreover, it is  necessary to  account for the
rate of collisions which depends on the relative  velocity at the point
of impact as well as all possibilities of collision  by summing up on the
locations of the impact on the sphere. $P_1(z)$ is then given by
\begin{align}\label{eq:7}
P_1(z)&=C\int_S\int\int d {\bf n} d{\bf u} d{\bf v} |({\bf v}-{\bf u}).{\bf n}| f({\bf v})\nonumber\\
&f_B({\bf u})\delta(z-{\bf v}.{\bf v^*}),
\end{align}
where  $f(v)$ and $f_B({\bf  u})$ are, respectively, the velocity distributions of
the tagged and bath  particles. The integral with   the
subscript $S$ corresponds to the integration  over the unit sphere with the
restriction $({\bf u}-{\bf v}).{\bf  n}<0$  where ${\bf n}$ is a unit vector along the axis joining
the two centers of particles (this imposes that the particles are approaching each other before colliding). $C$ is the normalization
constant such that $\int_{-\infty}^{\infty}dz P_1(z)=1$.

The postcollisional velocity ${\bf v^*}$ is given by the collision rule which is
\begin{equation}\label{eq:8}
{\bf v}^*={\bf v}+m\frac{1+\alpha}{m+M}[({\bf u}-{\bf v}).{\bf n}]{\bf n},
\end{equation}
where $M$ and $m$ are, respectively, the  mass of the tagged
and of the bath particle . $\alpha$ is the normal  restitution
coefficient comprised between $0$ and $1$.  For convenience\cite{Puglisi2006,SD06,PTV06},  we introduce $\alpha'$ such that
\begin{equation}
\frac{1+\alpha'}{2}=m\frac{1+\alpha}{M+m}.
\end{equation}

\subsection{One dimension}
In one dimension, the integral over angles is replaced by counting
the right and left collisions. Therefore, Eq.(\ref{eq:7}) becomes

\begin{equation}\label{eq:9}
P^{(1)}_1(z)=C\int\int d u d v | u- v| f( v)f_B(u)\delta(z- vv^*).
\end{equation}
For granular gases,  even if the  velocity distributions of the tagged
particle and of the bath particle are Gaussian\cite{MP99,Garzo1999} (or close to
the Gaussian profile\cite{Brey2005b}),  the granular temperatures of these two species are always different
when $\alpha<1$. Let  us denote $\gamma$ the ratio
between the bath and the tagged particle temperatures. The velocity distribution of the two species read :

\begin{equation}\label{eq:10}
f(v)=\sqrt{\frac{M}{2\pi T}}e^{-\frac{Mv^2}{2\gamma T}}
\end{equation}
and 
\begin{equation}\label{eq:11}
f_B(v)=\sqrt{\frac{m}{2\pi T}}e^{-\frac{mv^2}{2 T}}.
\end{equation}
It is necessary to distinguish the case $z<0$, where the distribution $P^{(1)}_1(z)$ is given by
\begin{align}\label{eq:12}
P^{(1)}_1(z)&=\frac{4C}{(1+\alpha')^2 }\int dv\left(1-\frac{z}{v^2}\right)f(v)\times\nonumber\\
&\times f_B\left(
\frac{v(1-\alpha')}{1+\alpha'}-\frac{2z}{(1+\alpha')v}\right)
\end{align}
from the case $z>0$, where $P^{(1)}_1(z)$ is equal to
\begin{align}\label{eq:13}
P^{(1)}_1(z)=&\frac{4C}{(1+\alpha')^2  }\left[\int_0^{\sqrt{z}} dv-\int_{\sqrt{z}}^{+\infty}\right]\left(\frac{z}{v^2}-1\right)f(v)\times\nonumber\\
&\times f_B\left(
\frac{v(1-\alpha')}{1+\alpha'}-\frac{2z}{(1+\alpha')v}\right).
\end{align}

Explicit  integration   over the velocity can   be  performed and some
details of the calculation are given in Appendix A. Let us introduce
\begin{align}\label{eq:14}
a&=\sqrt{\frac{M}{\gamma T}+\frac{m}{T}\left(\frac{1-\alpha'}{1+\alpha'}\right)^2},\\\label{eq:15}
b&=\sqrt{\frac{m}{T}\left(\frac{2}{1+\alpha'}\right)^2},\\
c&=\frac{2m}{T}\frac{1-\alpha'}{(1+\alpha')^2}\label{eq:16}.
\end{align}
the first-collision velocity-correlation distribution  $P^{(1)}_1(z)$  then reads
for $z<0$
\begin{equation}\label{eq:17}
P^{(1)}_1(z)=P^{(1)}_1(0)e^{(ab+c)z}
\end{equation}
and for $z>0$
\begin{align}\label{eq:18}
P^{(1)}_1(z)&=P^{(1)}_1(0)e^{cz}\left(\frac{a-b}{a+b}e^{-abz}erf\left(\frac{a-b}{\sqrt{2}}\sqrt{z}\right)\right.\nonumber\\
&\left.+e^{ab z}erfc\left(\frac{a+b}{\sqrt{2}}\sqrt{z}\right)\right),
\end{align}
where $P^{(1)}_1(0)$ is given by 
\begin{equation}
P^{(1)}_1(0)=\frac{(a+b)(a^2b^2-c^2)}{2ab\sqrt{a^2+b^2-2c}}.
\end{equation}

Note that $P^{(1)}_1(z)$ is always
asymmetric,  contrary   to an uncorrelated  velocity  distribution; this
results from     the existence  of   correlations between     pre  and
post-collisional  velocities of  a  particle.  Secondly,  $P^{(1)}_1(z)$  is
finite when $z$ goes to $0$, which is not the case for $P^{(1)}_\infty(z)$.

To  simplify the above expressions, Eq.(\ref{eq:17})-(\ref{eq:18}) and
to allow us to discuss the physical results, we  now need to specify
the temperature ratio $\gamma$.  For inelastic  particles in a  polydisperse granular bath, $\gamma$
is in general a complicated function of parameters such as the bath composition, the heating mechanism and the coefficient of restitution. However, three interesting limiting cases
provide a simple   expression of the temperature  ratio :    (i)
monodisperse  system, for which $\gamma=1$  (in a Gaussian approximation);

(ii) a mixture of granular gases in the
limit  of infinite dilution. Martin and Piasecki\cite{MP99}  showed  that the velocity
distribution     of an inelastic tracer in an elastic bath    remains  Gaussian  with  a granular
temperature  of   the  tracer given  by   the relation $T_{eff}=\gamma
T_{bain}$ (equipartition does not hold); this ratio $\gamma$ is given by
\begin{equation}\label{eq:19}
\gamma = \frac{M}{m}\frac{1+\alpha'}{3-\alpha'}.
\end{equation}
(iii) Thermalized systems of  elastic  particles ($\alpha=1$),  $P^{(1)}_1(z)$ then provides a non  trivial information on the short-time dynamics even for equilibrium systems.

In the first case ($M=m$  and $\gamma=1$), $\alpha'=\alpha$.
This gives for the first-collision velocity-correlation distribution, for $z<0$
\begin{align}\label{eq:20}
P^{(1)}_1(z)=\frac{M}{T}\frac{2+\sqrt{2(1+\alpha^2)}}{2(1+\alpha)\sqrt{1+\alpha^2}}
e^{\frac{2}{(1+\alpha)^2}(\sqrt{2(1+\alpha^2)}+1-\alpha)\frac{Mz}{T}}
\end{align}
and for $z>0$,
\begin{align}\label{eq:21}
&P^{(1)}_1(z)=\frac{M}{T}\frac{e^{\frac{2(1-\alpha)}{(1+\alpha)^2}\frac{Mz}{T}}}{2(1+\alpha)\sqrt{1+\alpha^2}}\nonumber\\
&\left[(2-\sqrt{2(1+\alpha^2)})
e^{-\frac{2}{(1+\alpha)^2}(\sqrt{2(1+\alpha^2)})\frac{Mz}{T}}\right.\nonumber\\
&
erf\left(\frac{\sqrt{2}-\sqrt{(1+\alpha^2)}}{1+\alpha}\sqrt{\frac{Mz}{T}}\right)+\nonumber\\
&
+(2+\sqrt{2(1+\alpha^2)}) e^{\frac{2}{(1+\alpha)^2}(\sqrt{2(1+\alpha^2)})\frac{Mz}{T}}\nonumber\\
&\left.
erfc\left(\frac{\sqrt{2}+\sqrt{(1+\alpha^2)}}{1+\alpha}\sqrt{\frac{Mz}{T}}\right)\right].
\end{align}

The decay of  $P^{(1)}_1(z)$ also   depends  on both  the temperature  on the
coefficient of restitution $\alpha$.

In   the second  case   (granular  tracer   in  a thermalized   bath),
substituting Eq.(\ref{eq:19}) in Eqs.(\ref{eq:14})-(\ref{eq:15}), one
obtains that $a=b=\sqrt{\frac{m}{T}}\frac{2}{1+\alpha'}$, which gives a simple expression
for $P^{(1)}_1(z)$. For $z<0$
\begin{equation}\label{eq:22}
P^{(1)}_1(z)=\frac{m}{T}\frac{3-\alpha'}{(1+\alpha')^{3/2}}e^{\frac{3-\alpha'}{(1+\alpha')^2}\frac{2mz}{T}}.
\end{equation}
and
for $z>0$
\begin{equation}\label{eq:23}
P^{(1)}_1(z)=\frac{m}{T}\frac{3-\alpha'}{(1+\alpha')^{3/2}}e^{\frac{2m}{T}\frac{3-\alpha'}{(1+\alpha')^2}z}erfc\left(\frac{2}{1+\alpha'}\sqrt{\frac{2mz}{T}}\right)
\end{equation}
In  order    to show that  the   local   environment influences the
first-collision velocity-correlation distribution, we reexpress $P^{(1)}_1(z)$ in terms
of the    temperature  of   the   tracer   $\gamma  T$.   Substituting
Eq.(\ref{eq:19}) in Eqs.(\ref{eq:22})-(\ref{eq:23}), one obtains that
for                $z>0$                       $P^{(1)}_1(z)=\frac{M}{\gamma
T}\frac{1}{\sqrt{1+\alpha'}}e^{\frac{1}{(1+\alpha')}\frac{2Mz}{\gamma
T}}$. Comparison    with   Eq.(\ref{eq:20}) shows  that    $P^{(1)}_1(z)$ is
sensitive to the heating procedure, the temperature of the tracer particle being the same in both cases.

Finally, for  elastic  hard  particles ($\alpha=1$),  the equipartition
holds, $\gamma=1$,   and    therefore  Eqs.(\ref{eq:14})-(\ref{eq:16}) become $a=b=(M+m)/(2\sqrt{mT})$ and  $c=(M^2-m^2)/(4mT)$, and  $P^{(1)}_1(z)$ reads for $z<0$
\begin{equation}\label{eq:37}
P^{(1)}_1(z)=\frac{M}{2T}\sqrt{1+\frac{M}{m}}e^{\frac{M(M+m)}{2mT}z}
\end{equation}
and for $z>0$
\begin{equation}\label{eq:38}
P^{(1)}_1(z)=\frac{M}{2T}\sqrt{1+\frac{M}{m}}e^{\frac{M(M+m)}{2mT}z}erfc\left((M+m) \sqrt{\frac{z}{2mT}}\right).
\end{equation}
\begin{figure}
\centering
 \resizebox{8cm}{!}{\includegraphics{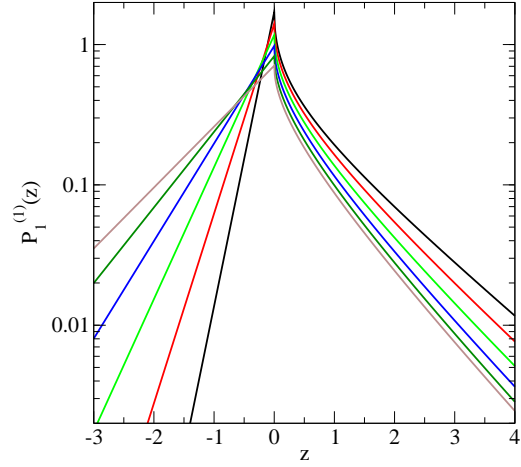}}
\caption{Log-linear plot of $P^{(1)}_1(z)$ versus $z$ ($M/T$ is set to $1$) (Eqs~(\ref{eq:20})-~(\ref{eq:21})).: Left, from top to bottom, the coefficient of restitution $\alpha=0,0.2,0.4,0.6,0.8,1$.}\label{fig:1}
 \end{figure}

\begin{figure}
\centering
 \resizebox{8cm}{!}{\includegraphics{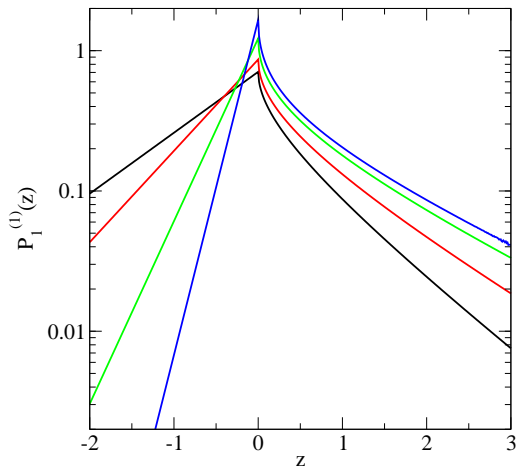}}
\caption{Log-linear plot of $P^{(1)}_1(z)$ versus $z$, for elastic hard spheres (Eqs.(\ref{eq:37})-~(\ref{eq:38})). The mass ratios are  $\frac{m}{M}=1,\frac{1}{2},\frac{1}{5},\frac{1}{10}$ (on the left side, from top curve to bottom curve).}\label{fig:2}
 \end{figure}

Figure~\ref{fig:1}  displays $P^{(1)}_1(z)$ as  a function  of $z$ for
different    values     of   the      coefficient  of      restitution
$\alpha=0,0.2,0.4,0.6,0.8,1$         for a monodisperse         system
(Eqs.(\ref{eq:20})-~(\ref{eq:21})).     For  elastic  hard  particles,
$P^{(1)}_1(z)$ is plotted in Fig.\ref{fig:2} as  a function of $z$ for
different values of the mass ratio $m/M=1,1/2,1/5,1/10$

Averaged quantities can be deduced from the first-collision distributions:
The integral of $P^{(1)}_1(z)$  over  $z<0$ corresponds  to the fraction  of
events in which the particle velocity after  collision is opposite to the
precollisional  velocity.  Integrating Eq.(\ref{eq:17}) over  all
negatives values of $z$ leads to
\begin{equation}
I_{P^{(1)}_1(z<0)}=\frac{(M_2+\sqrt{\gamma}(m+M))(M_2-\sqrt{\gamma}(M-m\alpha))}
{(1+\alpha)M_2\sqrt{mM+\gamma m^2}}
\end{equation}
with
\begin{equation}
M_2=\sqrt{(1+\alpha)^2mM+\gamma(M-m\alpha)^2}
\end{equation}

For elastic particles  (i.e.  $\alpha=1$), the fraction of collisions
in which the post-collisional velocity  has a direction opposite to that of the
 pre-collisional     velocity  becomes   very     simple  because  the
 equipartition property is satisfied ($\gamma=1$) :
\begin{equation}
I_{P^{(1)}_1(z<0)}=\sqrt{\frac{m}{m+M}}.
\end{equation}
Thus, for  a monodisperse system,  the probability of having a velocity
inverted after a collision is  higher than the  probability of  having a
velocity whose direction is not changed by the collision (in 1D).

\begin{figure}
\centering
 \resizebox{8cm}{!}{\includegraphics{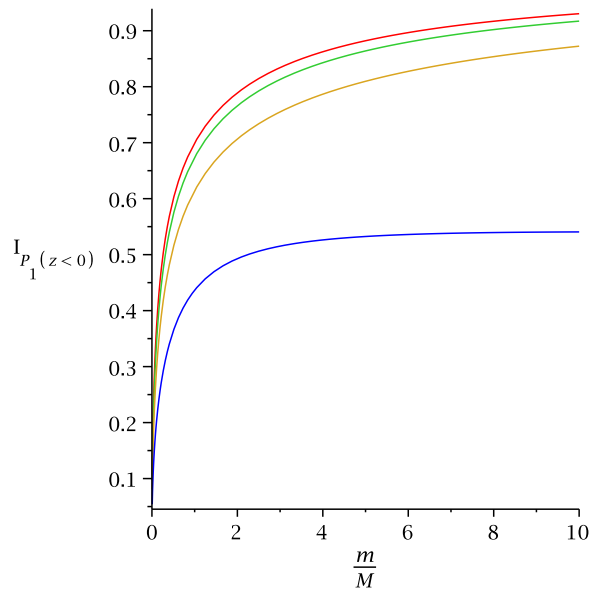}}
\caption{$I_{P^{(1)}_1(z<0)}$ versus mass ratio $m/M$  for different values of the coefficient of restitution $\alpha$: from top to bottom, $\alpha=0.99,0;8, 0.5, 0$.}\label{fig:3}
 \end{figure} 

For   inelastic  particles  ($\alpha<1$),   Fig.~\ref{fig:3}  shows that
$I_{P^{(1)}_1(z<0)}$  increases with the mass ratio $m/M$ for all  values  of $\alpha$. For
the sake of simplicity,  $\gamma$ is assumed equal  to $1$. Attention must be paid to the case $\alpha=0$: in this case, the limit
of $I_{P^{(1)}_1(z<0)}$  when   $m/M\rightarrow \infty$ is  equal  to  $1/2$
whereas  $I_{P^{(1)}_1(z<0)}$ goes to   $1$ when $m/M\rightarrow  \infty$ if
$\alpha>0$.   This      discontinuity   is    clearly    apparent   in
Fig.\ref{fig:3}.

\subsection{Two dimensions}
The  first-collision   velocity-correlation  distribution can   also  be obtained
analytically in two  dimensions and above.  Indeed the calculation can
be   performed  following  a method    used  in  the one-dimensional
case.     Let us note    that the   scalar   product  of the pre-  and
post-collisional velocities ${\bf v}.{\bf v}^*$  can be expressed in  an
orthonormal basis associated with the collision as
\begin{align}\label{eq:24}
{\bf v}.{\bf v^*}&=v_{n}v_{n}^*+v_{t}v_{t}^*,
\end{align}
where $v_n$ and $v_t$ denote  the normal and tangential components  of
the  velocity.     The  post-collisional quantities  in    the  rhs of
Eq.(\ref{eq:24}) can be eliminated by using Eq.(\ref{eq:8}):
\begin{equation}
{\bf v}.{\bf v^*}=v_t^2+\frac{1-\alpha'}{2}v_n^2+\frac{1+\alpha'}{2}v_nu_n,
\end{equation}
where $u_n$ is the normal component of the velocity of the bath particle.

Since the  normal and  tangent  components of the  tracer particle are
independent   random  variables,  the   integrals can  be 
performed successively over  $u_n$ and   $v_n$ as  in  the one-dimensional case,
provided that $z$ is shifted to $z-v_t^2$. It is worth  noting that the last integration over
angular variables (${\bf n}$) is trivial, since the integrand does not
depend on ${\bf n}$.

For  $z<0$, the decay of $P^{(2)}_1(z)$ remains exponential and is equal to

\begin{equation}
P^{(2)}_1(z) = P^{(1)}_1(0) \frac{ e^{(ab+c)z}}{\sqrt{1+2(ab+c)\frac{\gamma T}{M}}}.
\end{equation}

For $z>0$, $P^{(2)}_1(z)$ has two contributions,

\begin{align}
&P^{(2)}_1(z) =\frac{P^{(1)}_1(0) e^{(ab+c)z}}{\sqrt{1+2(ab+c)\frac{\gamma T}{M}}}erfc\left[\sqrt{z\left(\frac{M}{2\gamma T}+(ab+c)\right)}\right]\nonumber\\
&+P^{(1)}_1(0)\sqrt{\frac{M}{2\pi \gamma T}}e^{c z}\int_{0}^{z}dy\frac{1}{\sqrt{y}} e^{-(\frac{M }{2\gamma T}+c)y}\nonumber\\
&\left[e^{ab(z-y)}erfc\left[\frac{(a+b)\sqrt{(z-y)}} {\sqrt{2}} \right]\right.\nonumber\\
&\left.+\frac{a-b}{a+b} e^{-ab(z-y)}erf\left[\frac{(a-b)\sqrt{(z-y)}} {\sqrt{2}} \right]\right].
\end{align}

Note that $P^{(2)}_1(z)$ has an exponential decay when the scalar product of
velocities is negative (i.e. $z<0$),  with  the same coefficient $(ab+c)$  than  we
have obtained  in one dimension. The integration
over $v_t$ only changes the normalization constant compared to the  one-dimensional case.  Conversely,  for positive arguments
$z$, the shape of $P^{(2)}_1(z)$ is  more complicated than in one dimension.
As will be seen in the next section, the  exponential decay of $P^{(2)}_1(z)$ is
universal  since this result is   the same in  any  dimension and for
different heating mechanisms (through the $\gamma$ dependence). The fraction of events in which the scalar
product between the pre and post-collisional velocities is negative is
then given by 
\begin{equation}
I_{P^{(2)}_1(z<0)}=\frac{(a+b)(ab  -
c)}{2ab\sqrt{a^2+b^2-2c}} \frac{1}{\sqrt{1+2(ab+c)\frac{\gamma T}{M}}},
\end{equation}
where $a$, $b$ and $c$ are given by Eqs.(\ref{eq:14})-(\ref{eq:16}).

For instance, for elastic hard particles, the fraction of events in which the scalar product is negative after a collision is given by
\begin{equation}
I_{P^{(2)}_1(z<0)}=\sqrt{\frac{m}{m+M}}\sqrt{\frac{m}{M+2m}}.
\end{equation}

Therefore,        for      a       monodisperse    system     ($m=M$),
$I_{P^{(2)}_1(z<0)}=1/\sqrt{6}\simeq   0.40824..$     in     $2D$    whereas
$I_{P^{(1)}_1(z<0)}=1/\sqrt{2}\simeq 0.707..$  in  $1D$. In other words,  in
$2D$,  most collisions do not   change the scalar product,
whereas the converse   is observed in $1D$.

\subsection{Three dimensions}
In three dimensions,  the first-collision velocity-correlation distribution can be
similarly  derived.    The  scalar  product    between   the  pre-  and
post-collisional velocities can be expressed as

\begin{align}
{\bf v}.{\bf v^*}&=v_{n}v_{n}^*+v_{t}v_{t}^*+v_{z}v_{z}^*\nonumber\\
&=\frac{1-\alpha'}{2}v_n^2+\frac{1+\alpha'}{2}v_nu_n+v_{t}^2+v_{z}^2,
\end{align}
where $u_n$ is  the   normal component  of  the velocity  of the  bath
particle. $P^{(3)}_1(z)$  is  the probability distributions associated  with the
sum of two independent random variables. The first one is
the 1D collision term $\frac{1-\alpha'}{2}v_n^2+\frac{1+\alpha'}{2}v_nu_n$  and the second one is
the sum $v_{1t}^2+v_{1z}^2$, which is a $\chi^2_2$-distributed variable, and, as a result, an  exponentially
distributed variable :

\begin{equation}
\chi^2_2(y)=\frac{M}{2\gamma T}e^{-\frac{M y}{2\gamma T}}.
\end{equation}

$P^{(3)}_1(z)$  can be expressed as the convolution product of   the  two probability distributions of these variables. As the distributions are normalized, the distribution obtained via the convolution product is normalized. For $z<0$,
\begin{align}
P^{(3)}_1(z) =&P^{(1)}_1(0) \frac{M}{2\gamma T}\int_{0}^{+\infty}dy e^{-\frac{M y}{2 \gamma T}} e^{(ab+c) (z-y)}\nonumber\\\label{eq:25}
 =& P^{(1)}_1(0) \frac{ e^{(ab+c) z}}{1+2(ab+c)\frac{\gamma T}{M}}
\end{align}
and for $z>0$, $P^{(3)}_1(z)$ is the sum of several contributions :
\begin{align}\label{eq:26}
&P^{(3)}_1(z)=P^{(1)}_1(0) \left[\frac{M }{\gamma T}\frac{e^{(ab+c)z}erfc\left[\frac{(a+b)\sqrt{z}}{\sqrt{2}}\right]}{\frac{M}{\gamma T}+2(ab+c)}\right.\nonumber\\
&+\frac{M }{\gamma T}\frac{(b-a)e^{(-ab+c)z}erf\left[\frac{(a-b)\sqrt{z}}{\sqrt{2}}\right]}{(a+b)(-\frac{M}{\gamma T}+2ab-2c)}\nonumber\\
&\left.+\frac{M }{\gamma T}\frac{4abe^{-\frac{Mz}{2\gamma T}}\sqrt{a^2+b^2-2c-\frac{M}{\gamma T}}
erf\left[\frac{\sqrt{z}\sqrt{a^2+b^2-2c-\frac{M}{\gamma T}}}{\sqrt{2}}\right]}{(a+b)(-\frac{M}{\gamma T}+2ab-2c)(\frac{M}{\gamma T}+2(ab+c))}\right],
\end{align}
where $a$, $b$, and $c$ are given by Eq.(\ref{eq:14})-(\ref{eq:16}) and $\gamma$ is the temperature ratio.

As already noted, $P^{(3)}_1(z)$  has  the same $z$ dependence as in 1D and 2D when $z<0$. The influence of the dimension is in  the amplitude factor which decreases when  the dimension increases, the temperature
and other microscopic parameters (masses, coefficient of restitution)
being kept constant.

From Eq.(\ref{eq:25}), the fraction of events with a negative velocities scalar product can be exactly obtained; moreover,  a general formula can be obtained in any dimension:

\begin{equation}
I_{P^{(d)}_1(z<0)}=\frac{(a+b)(ab-c)}{2ab\sqrt{a^2+b^2-2c}}\left(1+2(ab+c)\frac{\gamma T}{M}\right)^{\frac{1-d}{2}}.
\end{equation}

\section{Influence of  the Sonine corrections of the velocity distribution on $P^{(1)}_1(z)$}
For  granular  gases,  whose  kinetic properties   are  well
described by the Boltzmann  equation, the velocity distribution is  no
longer a Gaussian. The deviations from  Gaussian behavior can be captured
by  Sonine corrections. It is  then possible to quantify the influence
of   these   corrections   on   the distribution
$P_1(z)$. (since the definition of the latter does  not depend on the
details of the velocity distribution function).

The non-linear Boltzmann equation can be solved numerically by using a
Direct Simulation  Monte  Carlo (DSMC) method\cite{B94,MS00}.  
We  have  performed DSMC simulations   for
monodisperse homogeneous  systems   excited  through a
stochastic thermostat. This allowed  us to compare the
distribution $P_1(z)$ obtained  by  simulation with the theoretical  prediction
calculated  with a Sonine correction.

The  calculation is similar to that  of  Eq.(\ref{eq:12}), but in the case
studied  here we  no longer consider  a tracer particle,   so that
$f=f_B$. Calculations were performed by considering only the first correction. This gives :

\begin{equation}
f({\bf v})=\left(\frac{M}{2\pi T}\right)^{d/2}e^{-\frac{M{\bf v}^2}{2T}}\left(1+a_2(\alpha)S_2\left(\frac{M{\bf v}^2}{2T}\right)\right),
\end{equation}

where $S_2$ is the second Sonine polynomial
expressed in the appropriate dimension. The  value of $a_2(\alpha)$ is
taken  equal to  its    usual  approximation for the stochastic thermostat\cite{VE98,MS00}.
 As the form  of   the
perturbation introduced is   simply a  multiplicative  polynomial, the
expressions  obtained for   the  1D  distribution $P^{(1)}_1(z)$ are
analytical.    As the result for  $z>0$   is involved, we will only
provide that obtained for $z<0$, which is

\begin{align}
&P^{(1)}_1(z)=\frac{e^{\frac{Mz}{T}\frac{2\left(1-\alpha+\sqrt{2}\sqrt{1+\alpha^2}\right)}{(1+\alpha)^2}}} 
{64\sqrt{2\pi}\left(1+\alpha+\alpha^2 +\alpha^3\right)^3}\times\nonumber\\
&\times\left(Q(\alpha)+a_2(\alpha)\left(Q_0(\alpha)+\frac{Mz}{T}Q_1(\alpha)+\frac{M^2z^2}{T^2}Q_2(\alpha)\right)\right)
\end{align}

where  the $Q_i(\alpha)$'s are simple   functions  of $\alpha$ given  in
Appendix  B.  

\begin{figure}
\centering
 \resizebox{9cm}{!}{\includegraphics{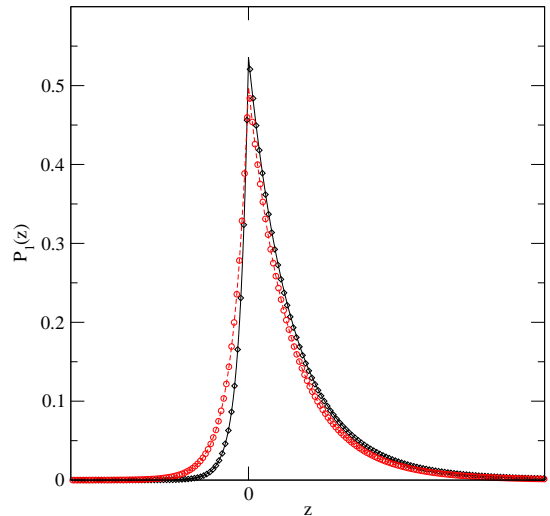}}
\caption{DSMC results(symbols) and theoretical predictions with the first Sonine correction(lines) for $P^{(2)}_1(z)$ versus $z=mv_1v'_1/T$ in 2D. Diamonds and circles correspond respectively to the simulations for $\alpha=0.2, 0.5$, respectively.}\label{fig:4}
 \end{figure}

Since  the 2D case  is closer to possible experimental systems,  we have  also included  Sonine  corrections to the
velocity  distribution  for   calculating the     velocity-correlation
distribution $P^{(2)}_1(z)$, but the lengthy expressions are not given
here.  Figure \ref{fig:4}  displays the  analytical result  ($P^{(2)}_1(z)$
with  Sonine corrections) and the  DSMC results  for  two  values of  the
coefficient of   restitution :  $\alpha=0.2,0.5$. Even    for the   more
inelastic  case, the agreement between  analytical results and DSMC is
remarkable. 

\section{Second and higher-collision velocity-correlation distributions}
The collision statistics  can be followed  beyond the first-collision
distribution. Formally, the second-collision  velocity-correlation distribution
$P_2(z)$ is given by the relation
\begin{align}
P_2(z)&=C_2\int_{S_1}\int_{S_2}\int\int d {\bf n}_1 d{\bf n}_2 d{\bf u}_1  d{\bf u}_2 d{\bf v} |({\bf u}_1-{\bf v}_1).{\bf n}_1|\nonumber\\
&|({\bf u}_2-{\bf v}_1^*).{\bf n}| f({\bf v})
f_B({\bf u}_1)f_B({\bf u}_2)\delta(z-{\bf v}.{\bf v}_2^*),
\end{align}

where ${\bf  v}_1$ denotes the pre-collisional  velocity of the tagged
particle for the  first collision, ${\bf  v}_1^*$ the velocity after the first collision and ${\bf v}_2^*$  the post-collisional  velocity after  the
second collision. ${\bf u}_1$ and ${\bf u}_2$ correspond to the velocities
of     the    bath particles    for  the    first and second collisions,
respectively. (Recall that $f({\bf  v})$ and $f_B({\bf v})$  are the velocity distributions of the tagged and bath particles, respectively). Obviously, ${\bf
v}_2={\bf  v}_1^*$, and by   using Eq.(\ref{eq:8}), the collision rule
gives for the two collisions
\begin{align}
{\bf v}_1^*&={\bf v}_1+\frac{1+\alpha'}{2}[({\bf u}_1-{\bf v}_1).{\bf n}_1]{\bf n}_1,\\
{\bf v}_2^*&={\bf v}_1^*+\frac{1+\alpha'}{2}[({\bf u}_2-{\bf v}_2).{\bf n}_2]{\bf n}_2.
\end{align}

Finally, $C_2$  is   the  normalization  constant  ensuring  that
\begin{equation}\label{eq:27}
\int_{-\infty}^{\infty} dz P^{(1)}_2(z)=1.
\end{equation}
In  a similar way,  it is possible  to write down closed equations for
$P_n(z)$.  However, whereas  tractable expressions have  been obtained
in any dimension for  $P_1(z)$, the calculation  increases drastically in complexity
for obtaining   the  distribution at  the  second
collision. In the restricted case  where $\alpha'=1$ in one dimension,
it   is  nonetheless  possible     to  obtain  the  exact   expression  of
$P^{(1)}_2(z)$ (details  of    the calulation are given   in  Appendix
C). Thus, for $z<0$,
\begin{align}\label{eq:28}
P^{(1)}_2(z)&=C'_2\left(\left(\frac{e^{\sqrt{2}abz}}{a}+\frac{e^{b\sqrt{a^2+b^2}z}}{b}\right) +\frac{(1+b^2z)}{b}\times\right.\nonumber\\
&\left.\times\int_0^\infty dv\frac{e^{-\frac{a^2v^2}{2}-\frac{b^2z^2}{2v^2}}}{v}\left[erfc\left(\frac{bv}{\sqrt{2}}\right)+erf\left(\frac{bz}{\sqrt{2}v}\right)\right]\right)
\end{align}
and for $z>0$,
\begin{align}\label{eq:29}
P^{(1)}_2(z)&=C'_2\left(\frac{\sqrt{2\pi}}{b^2}\left(\frac{e^{-\sqrt{2}abz}}{a} erf\left(\frac{(\sqrt{2}a-2b)\sqrt{z}}{2}\right)\right.\right.\nonumber\\
&+\frac{e^{\sqrt{2}abz}}{a} erfc\left(\frac{(\sqrt{2}a+2b)\sqrt{z}}{2}\right)\nonumber\\
&-\frac{e^{-b\sqrt{a^2+b^2}z}}{b}erfc\left(\frac{(-b+\sqrt{a^2+b^2})\sqrt{z}}{\sqrt{2}}\right)\nonumber\\
&+\frac{e^{b\sqrt{a^2+b^2}z}}{b}erfc\left(\frac{(b+\sqrt{a^2+b^2})\sqrt{z}}{\sqrt{2}}\right)\nonumber\\
&+2\frac{\sqrt{2\pi}(1+b^2z)}{b^3}\int_{\sqrt{z}}^{\infty} dv \frac{e^{-\frac{a^2v^2}{2}-\frac{b^2z^2}{2v^2}}}{v}\times\nonumber\\
&\left.\left[erfc\left(\frac{bv}{\sqrt{2}}\right)+erf\left(\frac{bz}{\sqrt{2}v}\right)\right]\right),
\end{align}
where $C'_2$ is determined from Eq.~(\ref{eq:27}) and, with the help of Eqs.(\ref{eq:14}) and (\ref{eq:15}), $a=\sqrt{\frac{M}{\gamma T}}$ and $b=\sqrt{\frac{m}{T}}$.
From Eqs.(\ref{eq:28}) and (\ref{eq:29}), one obtains the small-$z$ expansion for $P^{(1)}_2(z)$ :
\begin{equation}
P^{(1)}_2(z)\sim K_0(\sqrt{\frac{ab}{\gamma}}\frac{|z|}{T}).
\end{equation}
Therefore,     the second-collision velocity-correlation  distribution
shows a  divergence    at $z=0$, reminiscent  of   the   divergence of
$P^{(1)}_\infty(z)$,    when   the     two    velocities   are    completely
uncorrelated.  However, note  that   the coefficient of the   modified
Bessel function of the second  kind $K_0(z)$ is $ab$  instead of $a^2$ for
$P^{(1)}_\infty(z)$.  To  continue the analysis of the velocity-correlation
distributions in general, we   have performed DSMC in  various situations,
and monitored several $P^{(1)}_n(z)$'s.

\begin{figure}[t]
\centering
 \resizebox{8cm}{!}{\includegraphics{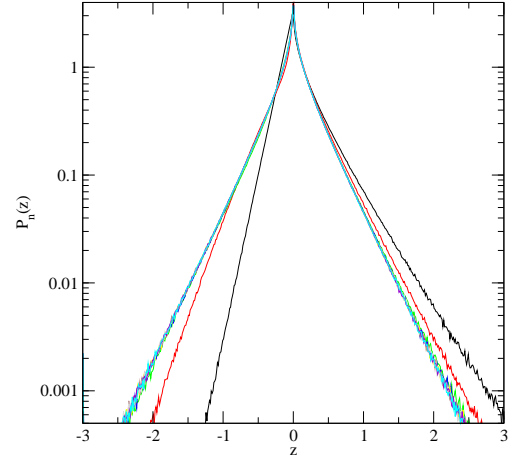}}
\caption{$nth$ collision velocity-correlation distributions $P^{(1)}_n(z)$ versus $z$ for various
 values of $n$: $n=1,2,..10$ ($D=1$). }\label{fig:5}
 \end{figure}

\begin{figure}[t]
\centering
 \resizebox{8cm}{!}{\includegraphics{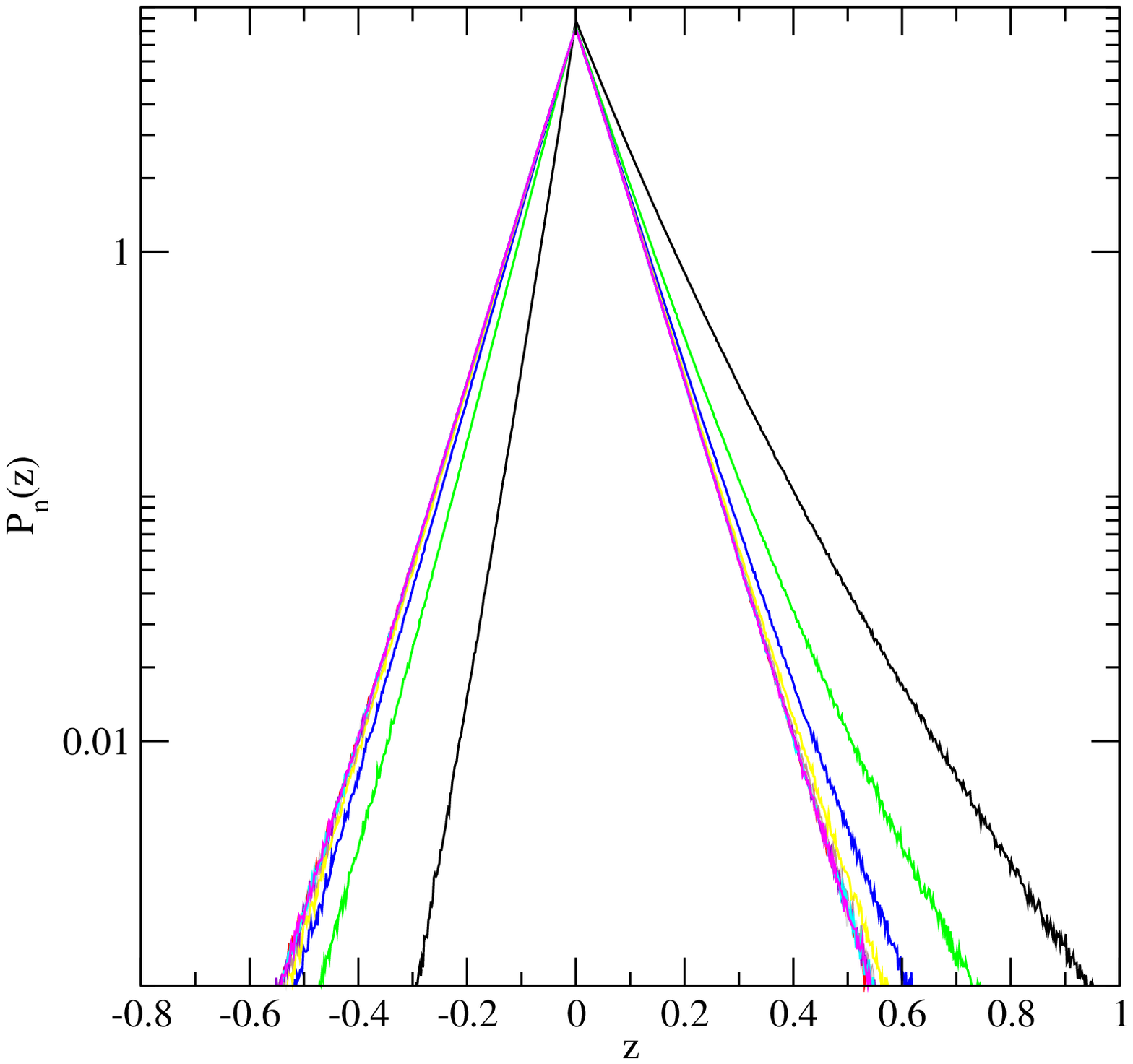}}
\caption{$nth$ collision velocity-correlation distributions $P^{(2)}_n(z)$ versus $z$ for various
 values of $n$: $n=1,2,..,10$ ($D=2$). }\label{fig:6}
\end{figure}

\begin{figure}[t]
\centering
 \resizebox{8cm}{!}{\includegraphics{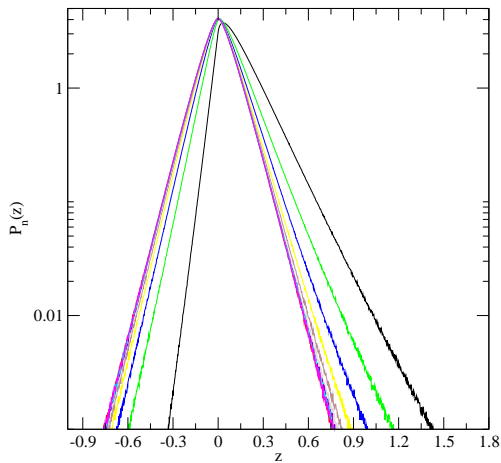}}
\caption{$nth$ collision velocity-correlation distributions $P^{(3)}_n(z)$ versus $z$ for various
 values of $n$: $n=1,2,..,10$ ($D=3$). }\label{fig:7}
 \end{figure}

 Figure    \ref{fig:5} shows  $P^{(1)}_n(z)$    as a  function  of  $z$  for
 $n=1,2..10$.  Note that  for    $n\geq 3$, the distribution  practically  reaches   the  asymptotic   value,  namely
 $P_\infty(z)$. A simple  physical  interpretation is that   after $3$
 collisions the   systems   loses  memory  of its     initial velocity
 configuration, and  the correlation between  the initial velocity and
 the velocity after  $n$ collisions vanishes when  $n$  is larger than
 $3$.  Similar plots are displayed  in Figs.~\ref{fig:6}
 and \ref{fig:7} for  $P^{(2)}_n(z)$ and $P^{(3)}_n(z)$,  respectively. One notes that  the
 convergence to the asymptotic  function, $P_\infty(z)$, becomes slower
 when the space dimension increases.

\section{Conclusion}
We have  introduced new velocity-correlation distributions  that
capture  the   early stages   of  the  dynamics.   These  non-trivial
quantities are efficient probes for investigating the environment of
a particle  in granular gases. We have  shown that these distributions
decay exponentially when the  scalar   product of the  velocities   is
negative. 

Other interesting physical situations could be considered: free cooling states, in which the velocity distribution obeys a scaling form,
binary mixtures\cite{BT02,BMP02,S03}, and not only in the infinite dilution limit considered in this paper. 

These distributions are easily accessible in computer simulations, and probably, in experiments; indeed, when the time resolution is smaller than the mean collision time, the probability that two collisions occur during a time step is small and the collision history can be monitored accurately, which permits to build the first-collision velocity-correlation distributions.
\section{Acknowledgments}
We thank Kristin Combs, Jeffrey Olafsen, Julian Talbot and Gilles Tarjus for suggestions and fruitful discussions.
\appendix
\section{1D first-collisions}

We have the following integrals, for $z<0$,
\begin{equation}
\int_0^{+\infty}dv(1-\frac{z}{v^2})\exp\left(-\frac{a^2v^2}{2}-\frac{b^2z^2}{2v^2}\right)=
\frac{a+b}{ab}\sqrt{\frac{\pi}{2}}e^{abz}
\end{equation}
 and for $z>0$,
\begin{align}
&\int_0^{\sqrt{z}}dv(\frac{z}{v^2}-1)\exp\left(-\frac{a^2v^2}{2}-\frac{b^2z^2}{2v^2}\right)=\nonumber\\
&\sqrt{\frac{\pi}{2}} \frac{1}{2 ab } \left((a-b) e^{-ab z}
\left(erf\left(\frac{a-b}{\sqrt{2}}\sqrt{z}\right)+1\right)+\right.\nonumber\\
&\left.
(a+b)e^{abz}
erfc\left(\frac{a+b}{\sqrt{2}}\sqrt{z}\right)\right)
\end{align}
and 
\begin{align}
&\int_{\sqrt{z}}^{\infty}dv (1-\frac{z}{v^2} )\exp\left(\frac{a^2v^2}{2}-\frac{b^2z^2}{2v^2}\right)=\nonumber\\
&\sqrt{\frac{\pi}{2}} \frac{1}{2ab } \left((a-b) e^{-abz}\left(
erf\left(\frac{a-b}{\sqrt{2}}\sqrt{z}\right)-1\right)+\right.\nonumber\\
&\left.
(a+b)e^{abz}
erfc\left(\frac{a+b}{\sqrt{2}}\sqrt{z}\right)\right),
\end{align}
which gives for $z<0$
\begin{equation}
P^{(1)}_1(z)=\frac{4C}{(1+\alpha')^2}\frac{a+b}{ab}e^{(ab+c)z}
\end{equation}
and for $z>0$
\begin{align}
P^{(1)}_1(z)&=\frac{4C}{(1+\alpha')^2}\sqrt{\frac{\pi}{2}}e^{cz}\left(\frac{a-b}{ab} e^{-abz}\left(
erf\left(\frac{a-b}{\sqrt{2}}\sqrt{z}\right)\right)\right.+\nonumber\\
&+\left. \frac{a+b}{ab}e^{abz}
erfc\left(\frac{a+b}{\sqrt{2}}\sqrt{z}\right)\right).
\end{align}
The constant $C$ can be obtained by calculating the normalization condition $\int dz P_1(z)=1$, which gives
\begin{equation}
C=\frac{(1+\alpha')^2}{4}\frac{a^2b^2-c^2}{\sqrt{2\pi(a^2+b^2-2c)}}.
\end{equation}

\section{Sonine correction to the 1D calculation}

For $z<0$, one finds

\begin{align}\label{eq:30}
& P^{(1)}_1(z)=e^{\frac{2(1-\alpha)}{(1+\alpha)^2}\frac{Mz}{T}}C\int_{0}^{\infty}du(1-\frac{z}{u^2})\nonumber\\
&\exp\left(-\frac{M}{T}\left(\frac{(1+\alpha^2)u^2}{(1+\alpha)^2}+\frac{2 z^2}{u^2(1+\alpha)^2}\right)\right)\nonumber\\
&\left(1+a_2(\alpha)\left(S_2\left(\frac{Mu^2}{2T}\right)+
S_2\left(\frac{M u^2}{T}\frac{(2z/u^2-(1-\alpha))^2}{2(1+\alpha)^2}\right)\right)\right).
\end{align}

After integration of Eq.(\ref{eq:30}), one obtains
\begin{align}
P^{(1)}_1(z)&=\frac{P_1(0) } {Q(\alpha)+a_2(\alpha)Q_0(\alpha)}e^{\frac{Mz}{T}\frac{2\left(1-\alpha+\sqrt{2}\sqrt{1+\alpha^2}\right)}{(1+\alpha)^2}}\nonumber\\
&\left(Q(\alpha)+a_2(\alpha)\left(Q_0(\alpha)+\frac{Mz}{T}Q_1(\alpha)+\frac{M^2z^2}{T^2}Q_2(\alpha)\right)\right)
\end{align}

where
\begin{align}
 Q(\alpha)=&16(1+\alpha)^4\left(1+\alpha^2\right)^2\nonumber\\
&\left(1+\alpha^2+\sqrt{2}\sqrt{1+\alpha^2}\right),
\end{align}

\begin{align}
Q_0(\alpha)&=3(1+\alpha)^4\nonumber\\
&\left(2+6\alpha^2+6\alpha^4+2\alpha^6+
\sqrt{2}\sqrt{1+\alpha^2}\left(1+6\alpha^2+\alpha^4\right)\right),
\end{align}

\begin{align}
&Q_1(\alpha)=2\sqrt{2}(1+\alpha)^2\left(1+\alpha^2\right)^{3/2}\nonumber\\
& (37-12\alpha+42\alpha^2-20\alpha^3+17\alpha^4)\nonumber\\
&+2(1+\alpha)^2\left(1+\alpha^2\right)(54-16\alpha\nonumber\\
&+84\alpha^2-32\alpha^3+54\alpha^4-16\alpha^5),
\end{align}

\begin{align}
&Q_2(\alpha)=8\left(1+\alpha^2\right)(3-2\alpha+3\alpha^2)\nonumber\\
&\left((7-6\alpha+3\alpha^2) \left (1+\alpha^2\right)+\sqrt{2}\sqrt{1+\alpha^2}\right.\nonumber\\
&\left.\left(5-4\alpha+5\alpha^2-2\alpha^3\right)\right).
\end{align}
\section{Second-collision velocity-correlation distribution}

In  one  dimension  when $\alpha'=1$,  the velocities   of  the tagged
particle  and     the  bath  particle   are   exchanged    during  the
collision. This   drastically   simplifies  the  expression  of    the
second-collision  velocity-collision distribution and the  calculation
becomes tractable. Indeed, if $\alpha'=1$, $P^{(1)}_2(z)$ becomes
\begin{align}\label{eq:31}
P^{(1)}_2(z)=C_2&\int dv\int du_1\int du_2 |u_1-v||u_2-u_1|f_B(u_1)\times\nonumber\\
&\times f_B(u_2)f(v)\delta(vu_2-z),
\end{align}
where $f_B(u)$ denotes the bath  velocity distribution and $f(v)$  the
tagged particle velocity distribution. 

We first integrate on the velocity of the bath particle $2$, namely the velocity of the colliding particle at the second collision. We drop the subscript of the velocity of the bath particle for the collision $1$, and $P^{(1)}_2(z)$  reads
\begin{equation}\label{eq:32}
P^{(1)}_2(z)=C_2\int                          dv\int                      du
|u-v||\frac{z}{v^2}-u|f_B(u)f_B\left(\frac{z}{v^2}\right) f(v).
\end{equation}

Let us introduce the function $I(z,v)$ :
\begin{equation}\label{eq:33}
I(z,v)=\int du |u-v||z-uv|e^{-\frac{b^2u^2}{2}}.
\end{equation}
When $z<v^2$, one has
\begin{align}\label{eq:34}
I(z,v)&=\frac{2v^2}{b^2}e^{-\frac{b^2z^2}{2v^2}}-\frac{2z}{b^2}e^{-\frac{b^2v^2}{2}}+\frac{\sqrt{2\pi}|v|(1+b^2z)}{b^3}\nonumber\\
&+\frac{\sqrt{2\pi}v(1+b^2z)}{b^3}\left(erf\left(\frac{bz}{\sqrt{2v}}\right)-erf\left(\frac{bv}{\sqrt{2}}\right)\right).
\end{align}
and when $z>v^2$
\begin{align}\label{eq:35}
I(z,v)&=-\frac{2v^2}{b^2}e^{-\frac{b^2z^2}{2v^2}}+\frac{2z}{b^2}e^{-\frac{b^2v^2}{2}}-\frac{\sqrt{2\pi}|v|(1+b^2z)}{b^3}\nonumber\\
&-\frac{\sqrt{2\pi}v(1+b^2z)}{b^3}\left(erf\left(\frac{bz}{\sqrt{2v}}\right)-erf\left(\frac{bv}{\sqrt{2}}\right)\right)
\end{align}
Inserting  Eq.(\ref{eq:34}) in Eq. (\ref{eq:31}) and integrating out the first two terms of the integrand leads to Eq.(\ref{eq:28}). 
For $z>0$, by using the property of $I(v,z)$ (Eqs.(\ref{eq:34})-~(\ref{eq:35})), for $z>0$, $P_2(z)$
is expressed as
\begin{equation}\label{eq:36}
P^{(1)}_2(z)=C_2\left(\int_{\sqrt{z}}^\infty dv-\int^{\sqrt{z}}_0 dv\left(\frac{e^{-\frac{a^2v^2}{2}-\frac{b^2z^2}{2v^2}}}{v^2}I(v,z)\right)\right).
\end{equation}
Integrating out the first terms of the right-hand-side of Eq.(\ref{eq:36}) leads to Eq.(\ref{eq:29}).


\end{document}